\documentclass[universe,article,accept,pdftex,moreauthors]{Definitions/mdpi} 
\firstpage{1} 
\pubvolume{1}
\issuenum{1}
\articlenumber{0}
\pubyear{2026}
\copyrightyear{2026}
\externaleditor{~~~} 
\datereceived{12 December 2025} 
\daterevised{16 January 2026} 
\dateaccepted{21 January 2026} 
\datepublished{ } 

\usepackage{placeins}
\usepackage{float}

\Title{Properties of Polarized Radio Sources in the Wide Chandra Deep Field South from 2 to 4~GHz}

\newcommand{\orcidauthorA}{} 

\Author{Samantha
 Adams \orcidauthorA{0009-0003-6268-660X}
 $^{1,2,3}$, Mark Lacy $^{2,*}$
, Preshanth Jagannathan $^{4}$, Jose Afonso $^{5}$, William Nielsen Brandt $^{6,7,8}$, 
B. M.  Gaensler $^{9}$, Evanthia Hatziminaoglou $^{10,11.12}$, Anna Kapinska
 $^{4}$, Josh Marvil $^{4}$, Hugo Messias $^{13}$, Steve Myers $^{4}$, Ray Norris $^{(14,15)}$, Kristina Nyland $^{16}$, Wiphu Rujopakarn $^{17}$, Nick Seymour $^{18}$, Mattia Vaccari $^{19}$ and Rick White $^{20}$}

\AuthorNames{Samantha Adams, Mark Lacy, Preshanth Jagannathan, Jose Afonso, William Niel Brandt, B. M. Gaensler, Evanthia Hatziminaoglou, Anna Kapinska, Josh Marvil, Hugo Messias, Steve Myers, Ray Norris, Kristina Nyland, Wiphu Rujopakarn, Nick Seymour, Mattia Vaccari and Rick White}

\address{
$^{1}$ \quad Department of Physics, University of Connecticut Unit 3046, 196 Auditorium Road, Storrs, CT 06269, USA; zjd24001@uconn.edu\\
$^{2}$ \quad National Radio Astronomy Observatory, Charlottesville, VA 22903, USA; mlacy@nrao.edu 
\\
$^{3}$ \quad Department of Physics, University of Rhode Island, Kingston, RI 02881, USA\\
$^{4}$ \quad National Radio Astronomy Observatory, Socorro, NM 87801, USA; (P.J.);(A.K.);(J.M.);(S.M.)\\ .
$^{5}$ \quad Department of Physics, University of Lisbon, 1749-016 Lisboa, Portugal\\
$^{6}$ \quad Department of Astronomy and Astrophysics, The Pennsylvania State University, \mbox{University Park, PA 16802, USA}\\
$^{7}$ \quad Institute for Gravitation and the Cosmos, The Pennsylvania State University, University Park, PA 16802, USA\\
$^{8}$ \quad Department of Physics, 104 Davey Lab,
 The Pennsylvania State University, University Park, PA 16802, USA\\
$^{9}$ \quad  University of California, Santa Cruz, Division of Physical \& Biological Sciences,
 1156 High Street, \mbox{Santa Cruz, CA 95064, USA}\\
$^{10}$\quad European Southern Observatory, 85748 {Garching b. München}
, Germany\\
$^{11}$\quad Instituto de Astrof\'{i}sica de Canarias, 38205 La Laguna, {Tenerife}, Spain\\
$^{12}$ \quad Departamento de Astrof\'{i}sica,Universidad de La Laguna, 38206 La Laguna, Tenerife, Spain\\
$^{13}$\quad Joint ALMA Observatory, Alonso de C\'{o}rdova 3107, 763-0355, Santiago, Vitacura, Chile
, Chile\\
$^{14}$\quad CSIRO Space \& Astronomy, P.O. Box 76, Epping, NSW 1710, Australia\\
$^{15}$\quad School of Science, Western Sydney University, Penrith, NSW 2751, Australia\\
$^{16}$\quad US Naval Observatory, 3450 Massachusetts Avenue NW, Washington, DC 20392, USA\\
$^{17}$\quad Department of Physics, Chulalongkorn University, Pathumwan, Bangkok 10330, Thailand\\
$^{18}$\quad Curtin Institute of Radio Astronomy, Curtin University, Bently, WA 6102, Australia\\
$^{19}$\quad Department of Astronomy, University of Cape Town, Rondebosch 7701, South Africa\\
$^{20}$\quad Space Telescope Science Institute, Baltimore, MD 21218, USA
}

\corres{Correspondence: mlacy@nrao.edu}

\abstract{We present a study of the linear polarization properties of radio sources within the 10~deg$^2$.
 Wide \emph{Chandra} 
 Deep Field South (W-CDFS) in S-band (2--4~GHz). Our W-CDFS image has an angular resolution of 15~arcsec and a 1$\sigma$ RMS in Stokes $I$ of $\approx$50 $\upmu$Jy/beam. We detect 1920 distinct source components in Stokes $I$ and 175 in linear polarization. We examine the polarized source counts, Faraday Rotation measures, and fractional polarization of the sources in the survey. We show that sources with a total intensity above $\approx$10~mJy have a mean fractional polarization value of $\approx$3\% from modeling the polarized counts. We also calculate an estimate for the limit on the fractional polarization level of sources with a total intensity below 1~mJy (mostly star-forming galaxies) of $\stackrel{<}{_{\sim}}$3\% using stacking. The mean Faraday Rotation we measure is consistent with that due to the Milky Way. We also show that fractional polarization is correlated with in-band spectral index, consistent with a lower mean fractional polarization for the flat-spectrum population. In addition to characterizing the S-band polarization properties of sources in the W-CDFS, this study will be used to validate the shallower, but higher angular resolution S-band polarimetric information that the VLA Sky Survey will provide for the whole sky above Declination $-$40 degrees over the next few years.}


\begin{document}


\section{Introduction}\label{sec1}

Surveys over $\sim$10\,deg$^2$ fields provide a useful insight into the high redshift universe, whilst at the same time spanning a range of environments from isolated galaxies to moderately rich clusters~\citep{2020ApJ...889..185K}. The Legacy Survey of Space and Time (LSST) has pre-defined five such fields as ``Deep Drilling Fields'' (DDFs) that will be repeated with a shorter cadence than the main survey, and will result in a deeper co-added dataset to $r\approx$ 27. Radio data in these fields at a frequency $\approx$1~GHz will be contributed by the MeerKAT MIGHTEE survey~\citep{2016mks..confE...6J,2022MNRAS.509.2150H,2024MNRAS.528.2511T, 2025MNRAS.536.2187H} and the ASKAP EMU survey~\citep{2021PASA...38...46N}. At 3~GHz, a MeerKAT survey is planned, but only over the central regions~\citep{2016mks..confE...6J}. 

\begin{figure}[H]
    \includegraphics[scale=0.6,trim=0.5in 0.5in 0.0in 0.5in]{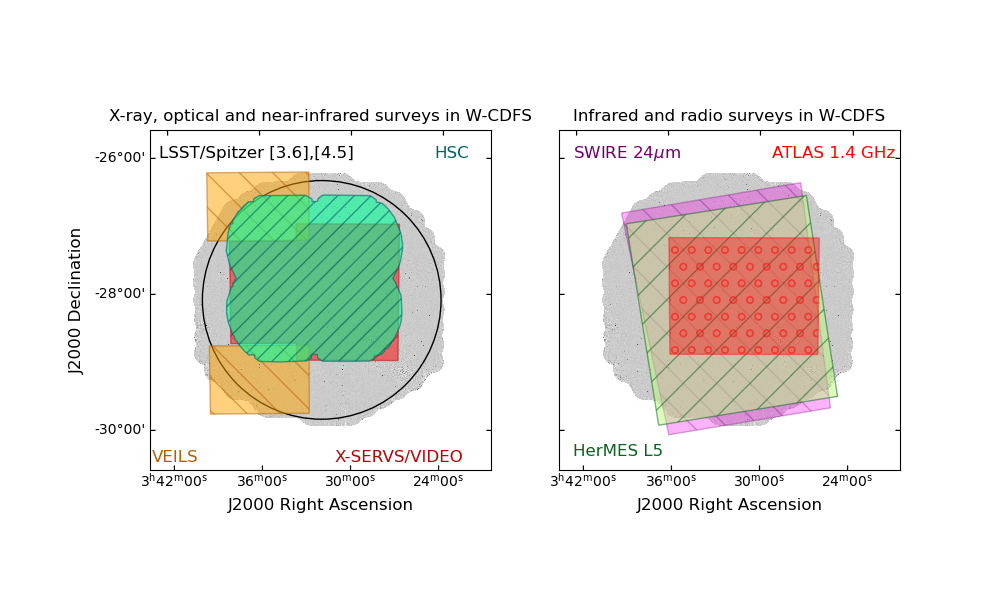}
    \caption{Surveys covering  $>$1 deg$^{2}$ in the W-CDFS, overlaid on the Stokes $I$ image from this paper (greyscale). \textbf{Left:}  X-ray through near-infrared. The \emph{XMM-Newton} XMM-SERVS survey~\citep{2021ApJS..256...21N} covers the footprint of the near-infrared VIDEO survey~\citep{2013MNRAS.428.1281J} in solid red. The LSST deep-drilling field is represented by the black circle, which is also the footprint covered by the warm \emph{Spitzer} 3.6 $\upmu$m and 4.5 $\upmu$m data~\citep{2021MNRAS.501..892L,2022A&A...658A.126E}. Further near-infrared data are provided by VEILS (orange, right-hatched~\citep{2017MNRAS.464.1693H}), which, together with VIDEO, covers the footprint of the Dark Energy Survey supernova fields in W-CDFS. W-CDFS also has deep optical imaging from the HyperSuprimeCam SSP survey (green, left-hatched~\citep{2018PASJ...70S...4A}). \textbf{Right:} Mid- and far-infrared and radio surveys, with the \emph{Spitzer} Wide-Area Infrared Extragalactic Survey (SWIRE) 24 $\upmu$m coverage (magenta, right-hatched~\citep{2003PASP..115..897L}), the \emph{Herschel} Multi-tiered Extragalactic Survey (HeRMES) (green, left-hatched~\citep{2012MNRAS.424.1614O}) and the 1.4~GHz component of the Australia Telescope Large Area Survey (ATLAS) (red circle speckled~\citep{2014MNRAS.441.2555H}). See \citet{2021MNRAS.501..892L} for more details and other surveys in this field.}\label{fig:wcdfs}
\end{figure}

This motivated us to image the entire 10~deg$^2$ of one of the DDFs, namely that centered on the \emph{Chandra} Deep Field South~\citep{2001ApJ...551..624G}, usually referred to as the Wide \emph{Chandra} Deep Field South (W-CDFS), e.g.,~\citep{2020A&A...634A..50P} (see Figure \ref{fig:wcdfs} for overlapping surveys), with the Karl G.\ Jansky Very Large Array (VLA) in S-band at 2--4~GHz at $\approx$~$15^{''}$ resolution as the beginning of a campaign to image the full extent of the three pre-defined DDFs observable by the telescope in S-band. In particular, we were interested in polarization properties of the sources, which have, to date, not been investigated for large numbers of faint ($\sim$1~mJy) radio sources at frequencies above 2~GHz. Radio synchrotron emission intrinsically has a high linear polarization (up to $\approx$70\% (e.g.,~\citep{1965ARA&A...3..297G}). The measured polarization from a cosmic radio source is typically much lower (only a few percent) as the intrinsic polarization fraction is reduced by line-of-sight effects, averaging within the observing beam and Faraday effects. To the extent that these can be disentangled, however, polarization provides a unique insight into the magnetic field structure and magnetionic environment of a radio source (\citep{1966MNRAS.133...67B,1991MNRAS.250..726T, 1998MNRAS.299..189S,2015aska.confE.103G}).

\textls[-15]{Above $\approx$1~mJy at 1.4~GHz ($\approx$0.5~mJy at 3~GHz for a steep-spectrum source), the radio source population is dominated by Active Galactic Nuclei (AGN) e.g.,
~\citep{Matthews_2021}.} {{Below} $\approx$1~mJy ($\approx$0.5~mJy at 3~GHz for a steep-spectrum source), star-forming galaxies, together with a small population of radio-quiet AGN, increasingly contribute to the radio source population in a redshift dependent way. So, a faint source survey will contain a mix of these populations~\citep{2016A&ARv..24...13P}.} Although the polarization properties of luminous AGN are well known, e.g.,~\citep{2016ApJ...825...59A,2017MNRAS.469.4034O,2018MNRAS.475.1736V}, the polarization properties of faint star-forming galaxies {{and radio-quiet AGN}} have not been so well studied. Nearby spiral galaxies can have a wide range of polarizations, from $\approx$1\% to up to $\approx$18\% for edge-on systems~\citep{2009ApJ...693.1392S}, but the polarization properties of star-forming objects at higher redshifts remain poorly constrained, with only one detection in an lensed galaxy~\citep{2023Natur.621..483G}. 
Nevertheless, these are important to understand, both from the point of view of studying the magnetic field in the interstellar medium (ISM) and its evolution with cosmic time, e.g.,~\citep{2004NewAR..48.1289B}, and for the possible use of polarization in radio weak lensing studies, where the magnetic field direction can give an indication of the intrinsic orientation of the galaxy, reducing the noise and systematics~\citep{2011MNRAS.410.2057B}.
Also, with respect to the AGN population, at $>$2~GHz, we might expect to see a higher fraction of flat-spectrum AGN than is typical in lower-frequency surveys and whose polarization properties are likely to be different from those of steep-spectrum AGN. 

Another important aspect of this survey was to provide validation data for the VLA Sky Survey (VLASS)~\citep{2020PASP..132c5001L}. VLASS is also an S-band survey, but at higher resolution (2.5$^{''}$) and less deep ($\approx$150 $\upmu$Jy per epoch for three epochs). We can thus use this lower-resolution survey to investigate the completeness of VLASS and to validate the polarization calibration for compact ($\stackrel{<}{_{\sim}}$~2$^{''}$) sources. 

Published radio surveys in CDFS are of smaller area: the Australia Telescope Large Area Survey (ATLAS)~\citep{2014MNRAS.440.3113H, 2015MNRAS.453.4020F, 2012A&A...544A..38Z} covers the central 3.6~deg$^2$ at 1.4 GHz and 2.3~GHz, with other ATLAS surveys covering the central $\approx$0.25~deg$^2$ at 5.5 and 9.0 GHz~\citep{2015MNRAS.454..952H,2020MNRAS.491.3395H}. In addition, the very center of the field, which contains the Hubble Ultra-Deep Field, has been the subject of deep imaging with the VLA at 6 and 3~GHz, covering $\approx$50 and 200 arcmin$^2$, respectively~\citep{2016ApJ...833...12R,2020ApJ...901..168A, Miller_2013}.


\section{Methods}\label{sec2}

\subsection{Observations}\label{sec2.1}
The 10 deg$^{2}$ of the LSST DDF in CDFS were observed with the VLA in C-configuration in S-band (2--4~GHz) as program 21A-017 between 2021-07-24.
 and 2021-09-09. The observations were split into 13 scheduling blocks, each of duration $\approx$2 h, that each observed a strip $\approx$200-arcmin in R.A. by 15-arcmin in Dec. The observations were made on a hexagonal grid of 513 pointings separated by 15~arcmin (two pointings [J2000 03:29:12.7--26:41:21 and 03:36:00.8--26:41:21] accidentally had their coordinates missed in one of the scheduling blocks and the subsequent pointings duplicated, resulting in two small ($\approx$0.1 deg$^2$ each) areas of uneven coverage; however, the pointing overlap is sufficient that the survey remains contiguous in those areas). The program was observed in filler time, with an intended coverage of three executions; however, the program was not completed, so only one complete coverage was obtained.
Each field was observed for approximately 110 s. The phase calibrator was J\,0402-3147. Flux density and bandpass calibration was performed using 3C\,138 (J~0521 + 1638) and the very low polarization object 3C\,84 (J\,0319 + 4130) was used as the polarization leakage calibrator.

\subsection{Calibrating and Imaging the Data} \label{sec:imaging}

Flux density, amplitude, and phase calibration were performed using the VLA pipeline version 2020.0.36 running under CASA version 6.1.2.7~\citep{2022PASP..134k4501C}. The pipeline does not yet perform polarization calibration, so this was carried out using the procedure outlined in the polarization calibration CASA guide (\url{https://casaguides.nrao.edu/index.php title=CASA_Guides:Polarization_Calibration_based_on_CASA_pipeline_standard_reduction:_The_radio_galaxy_3C75-CASA6.2.1} (accessed on 1 September 2021)).
 Cross-hand delays were solved per spectral window (``single-band'' delays), the leakage terms then derived assuming 3C~84 is unpolarized (``Df'' mode) and the $R-L$ phase calibrated using 3C~138 (\url{https://science.nrao.edu/facilities/vla/docs/manuals/obsguide/modes/pol\#section-4} (accessed on 1 September 2021)).  Based on the scatter in the calibration solutions, the on-axis polarization leakage is estimated to be 0.5\%, consistent with that expected for the VLA using an unpolarized leakage calibrator (\url{https://library.nrao.edu/public/memos/evla/EVLAM_177.pdf} (accessed on 1 September 2021)).
 
Imaging was performed in CASA. Each execution block was first imaged individually in CASA version 6.5.0.15 using uniform weighting and cleaned to 2~mJy within a mask that allowed cleaning of sources detected in VLASS Epoch 1 Quick Look images. The list of VLASS sources in the mask was generated using PyBDSF version 1.10.1~\citep{2015ascl.soft02007M}. The mask had 10-arcsec cleaning radius around faint ($<$10~mJy) VLASS sources and a 20-arcsec radius around brighter ($>$10~mJy) sources. The data were self-calibrated with a single iteration in phase only and a solution interval of 1020 s (the time interval between observations of the phase calibrators). The final Stokes $I$ imaging was carried out on the combined set of 13 execution blocks using the tclean task in the parallel mpicasa framework~\citep{2022ASPC..532..389E} with four CPUs in CASA version 6.6.4.34. The Stokes $I$ multifrequency synthesis mosaic image was made in CASA 6.5.0.15 with three Taylor terms and Briggs~\citep{1995PhDT.......238B} weighting with a robust $=$0.5. We used the mosaic gridder and cleaned down to 0.3~mJy/beam ($\approx$6$\sigma$). The RMS in source-free regions of the final image is $\approx$50 $\upmu$Jy/beam and the beam is $16.5 \times 11.0$-arcsec at a position angle of 8.3 degrees.

For the polarization images, we divided the 2~GHz bandwidth into four equal channels to reduce the effects of depolarization across the band, 
imaging in Stokes $IQUV$ using a H\"{o}gbom clean
~\citep{1974A&AS...15..417H}. The noise in each channel was approximately 120, 70, 70, and 120~$\upmu$Jy/beam
for the channels centered on 2.243, 2.755, 3.243, and 3.755~GHz, respectively (the outer channels suffered from worse radio frequency interference). 
As the fractional circular polarization of the extragalactic source population is known to be low ($<<$1\%~\citep{2000MNRAS.319..484R}), we restricted our analysis to linearly polarized flux density (Stokes $Q$ and $U$). Through the \textit{immath} command in CASA version 6.5.0.15, the linear polarization intensity was obtained for each channel and the channel images were then averaged together to create an image in polarized intensity. {{The} polarization of the primary beam of the VLA adds off-axis polarization leakage in addition to the on-axis term. This is difficult to quantify exactly; however, based on the limits from Stokes $I$ sources that are undetected in linear polarization in the image, this leakage  is $\stackrel{<}{_{\sim}}$2\% for Stokes $Q$ and $U$.}

PyBDSF~\citep{2015ascl.soft02007M} was used for source component detection in both the Stokes I and polarized intensity images, with a detection limit of $\approx$5$\sigma$.
In total, there were 1920 sources detected in the Stokes $I$ image, and 175 sources detected at $>$5$\sigma$ in the linearly polarized intensity image.{{The} catalog from the linearly polarized intensity image was used to identify polarized source components, but was not used for polarization measurements due to the non-Gaussian noise in the image. Instead,} the Stokes $Q$ and $U$ flux densities for each component were measured from the individual channel images as described in Section~\ref{sec:polcounts}.

We used \emph{Gaia} DR3 data~\citep{2023A&A...674A...1G} to check for positional accuracy. We found 132 matches within $\pm$1 arcsec of the radio component positions, with an RMS scatter of 0.4 arcsec in R.A. and 0.5 arcsec in Dec.\ and a median offset (radio---\emph{Gaia} position) of 0.0 $\pm$ 0.1 arcsec in R.A. and 0.3 $\pm$ 0.1 arcsec in Dec. We believe the scatter and the declination offset can probably be explained as a result of neglecting $w$-terms in the imaging (see VLASS memo 14, \url{https://library.nrao.edu/public/memos/vla/vlass/VLASS_014.pdf} (accessed on 1 December 2025)). As these values are small compared to the synthesized beam, we do not attempt any correction.

\section{Results}\label{sec3}
\subsection{Source Component Counts in Total Intensity}\label{sec3.1}

\begin{figure}[H]
     \includegraphics[width = 0.85\textwidth]{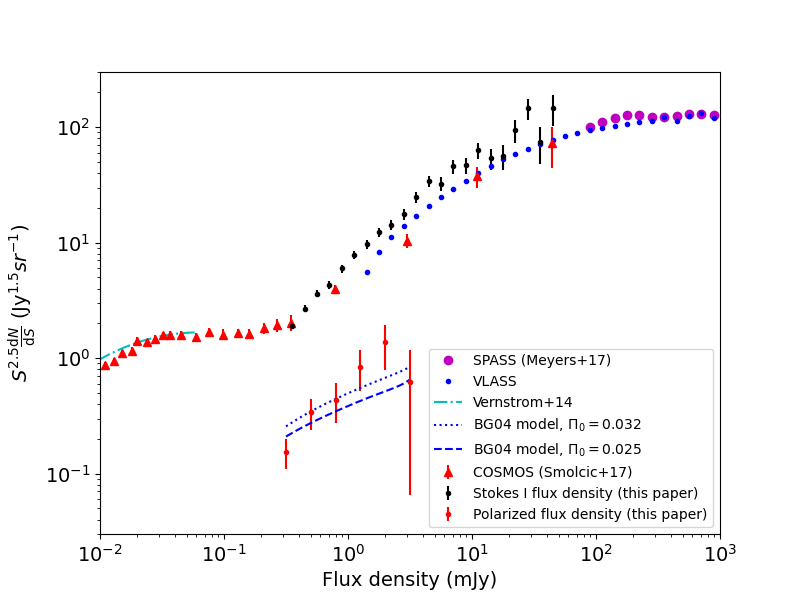}
     
%
%
%
     \includegraphics[width = 0.85\textwidth]{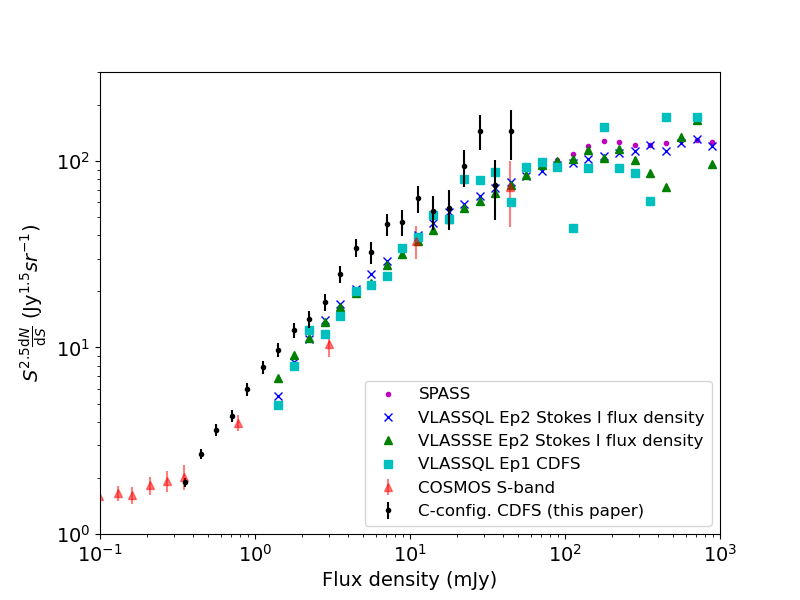}
     \caption{\textbf{(Upper panel)}
 counts of linear polarization and total intensity for the source components in the survey and in other $\approx$3~GHz surveys in the literature as detailed in Section~\ref{sec:imaging}. The full survey areas are used in all cases. The two models of the polarized counts correspond to the model of Equation \eqref{eqn:polcts}, with the dotted line having the characteristic polarization fraction that matches the 1.4~GHz population and the dashed one the best fit to our S-band data. \textbf{(Lower panel)} Detail of the total intensity counts, focusing on the region of overlap of the COSMOS, VLASS, and SPASS counts with those from this paper.}
     \label{fig:pol counts}
 \end{figure}

We plot the source counts in both total intensity and linearly polarized intensity in the upper panel of Figure \ref{fig:pol counts} (we have not attempted to combine components corresponding to the same physical source; however, only about 10\% of sources are resolved into multiple components at the angular resolution of our survey).
We have also included counts from the S-band Polarisation All-Sky Survey (S-PASS) at 2.3~GHz from \citet{2017PASA...34...13M} (from 16,600 deg$^2$ of sky to a flux density limit of $\approx$100~mJy, adjusted to 3.0~GHz by assuming a mean spectral index of $-$0.7), the counts from the VLASS Quick Look Epoch 2 images (from 34,000 deg$^2$ to $\approx$1 mJy, available from \url{https://cirada.ca/vlasscatalogueql0}, details of the similar Epoch 1 catalog in~\citep{2021ApJS..255...30G}), the counts from the S-COSMOS survey (from 2~deg$^2$ of sky to $\approx$10 $\upmu$Jy~\citep{2017A&A...602A...1S}) and ultra-deep counts (from 0.02~deg$^2$ of sky to $\approx$50 nJy \citet{2014MNRAS.440.2791V}). The three latter sets of counts are all at 3.0 GHz.
The VLASS counts are lower by about 40\% compared to the counts from our C-configuration survey,  but our counts are consistent with the S-COSMOS counts at faint fluxes. We therefore decided to investigate the cause of the apparent tension between the VLASS counts and our \mbox{C-configuration counts.}

 A deficit in the VLASS counts is to be expected based on the analysis of \citet{2015arXiv150205616C}, who predicted a $>$25\% incompleteness in the VLASS source counts due to VLASS resolving flux from extended sources.
However, the C-configuration counts are also higher than the VLASS counts at bright fluxes. Complicating this analysis is that half of VLASS Epoch 1 suffered from an issue with the antenna pointing, leading to larger flux density uncertainties and a possible bias (\url{https://library.nrao.edu/public/memos/vla/vlass/VLASS_013.pdf} (accessed on 1 December 2025)). In the lower panel of Figure \ref{fig:pol counts}, we have therefore plotted the counts for the surveys listed above (with the exception of the ultra-deep counts of \citet{2014MNRAS.440.2791V})  as well as several sets of counts from VLASS. These include 
the counts from the Canadian Initiative for Radio Astronomy Data Analysis (CIRADA) catalogs (\url{https://cirada.ca/vlasscatalogueql0} (accessed on 1 December 2025)) for Epoch 2 of the VLASS Quick Look images and the Single Epoch products from Epoch 2, that are self-calibrated and better cleaned than the Quick Look images (1130 deg$^2$ at the time of writing, though unfortunately with no overlap with the CDFS field). Finally, we also plot the VLASS Epoch 1 Quick Look counts from \citet{2021ApJS..255...30G}, but only in the same CDFS area as our survey (to investigate possible local over- or under-densities). The lower panel of Figure \ref{fig:pol counts} shows that the VLASS counts from different epochs and processings are all consistent with one another, and also indicates that the excess in counts in the CDFS field at $\approx$20~mJy is present in both the VLASS and C-configuration counts in the CDFS and is therefore most likely due to a local over-density of sources. 
At the bright end, the VLASS counts are only slightly below the SPASS counts, but at bright flux densities, the 3~GHz source population is increasingly dominated by unresolved flat-spectrum sources, so better agreement is expected.

\vspace{-10pt}

First, we checked for consistency in the flux density scales between VLASS and our CDFS survey by directly comparing the flux densities of compact ($<$1-arcsec), bright ($>$1~mJy) source components common to our survey, and VLASS Epoch 2 Quick Look (\mbox{147 objects}), finding a median flux density ratio of 0.95 $\pm$ 0.02, suggesting there is no significant issue with the flux-density scales, whose accuracy is expected to be $\approx$3--5\%~\citep{2017ApJS..230....7P}. We thus infer that the deficit in the VLASS counts is indeed due to resolved objects not having all their flux density included in the pyBDSF Gaussian component fits. 
Consistent with this, we find that 65\% of VLASS source components in the CDFS field are significantly resolved (deconvolved sizes $>$1-arcsec), and that source components that are slightly resolved in our C-configuration survey (with peak to total flux ratios $<$0.8, corresponding to 12\% of the sources) are, on average, 37\% brighter than their matches in the VLASS Epoch 2 Quick Look catalog.

\subsection{Source Component Counts in Polarized Intensity}\label{sec:polcounts}

For the polarized counts, we follow \cite{2014MNRAS.440.3113H},
who focused on the polarization properties of radio sources at 1.4 GHz. Their sample was observed at a similar depth and includes roughly the same number of sources as ours, covering the central 3.6 deg$^2$ of the W-CDFS and 2.8 deg$^2$ of the ELAIS-S1 field. At these polarized flux densities, the population is dominated by AGN.

To construct a catalog of the polarized source components, we used the pyBDSF component list from the polarized intensity image described in Section~\ref{sec:imaging}. We then took the component positions from this list and cataloged the pixel values in Stokes $I, Q$, and $U$ at the position of each component. The polarized flux for each channel was then estimated as $S_P=\sqrt{Q^2+U^2}$ and averaged to obtain a mean value of $S_P$ and fractional polarization, $\Pi = S_P/I$, across the band. Sources with Stokes $I$ peak flux densities $<$0.5~mJy/beam ($\approx$5$\sigma$) in any band were removed from the final catalog, which consists of 175 objects. 

\begin{figure}[H]
    \includegraphics[scale=0.4]{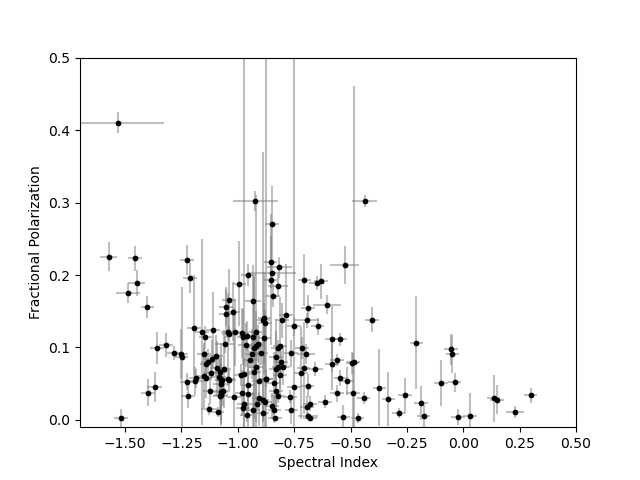}
    \caption{Fractional polarization versus in-band spectral index for the 175 sources in the polarized source catalog.}
    \label{fig:Pol_vs_alpha}
\end{figure}

There were some expected differences between our results and those of \cite{2014MNRAS.440.3113H} as our observations were made at a higher frequency, slightly reducing depolarization from Faraday Rotation, but also including a higher proportion of flat-spectrum AGN. We find that the polarized counts are quite well modeled (reduced $\chi^2=1.9$; probability of $\chi^2$ given the five degrees of freedom of 0.13) using the approach of \citet{2004NewAR..48.1289B} (hereafter BG04) for sources in the NVSS survey. We used the parameters for the BG04 model listed in \citet{2014MNRAS.440.3113H} with a fractional linear polarization distribution $f (\Pi)$:
\begin{equation}\label{eqn:polcts}
    f(\Pi)=\frac{a}{\Pi\, {\rm log}(10)} {\rm exp} \left\{ \frac{-\left[ {\rm log_{10}} (\Pi/\Pi_0) \right]^2}{2\sigma_p^2} \right\}
\end{equation}
with a fiducial fractional polarization $\Pi_0$ such that  ${\rm log_{10}} \Pi_0 = -1.5$,
\[a = \left\{ \begin{array}{ll}
   0.690  &{\rm if\,\, log_{10}}(\Pi) \leq -2  \\
   0.808  &{\rm if\,\, log_{10}}(\Pi) > -2\\
\end{array} \right\}
\]
and dispersion $\sigma_P$:
\[\sigma_P= \left\{ \begin{array}{ll}
   0.700  &{\rm if\,\, log_{10}}(\Pi) \leq -2  \\
   0.550  &{\rm if\,\, -2< log_{10}}(\Pi) \leq -1.5\\
   0.353  &{\rm if\,\, log_{10}}(\Pi)> -1.5\\
\end{array} \right\}
\]
with a mean $\bar{f}=0.033$. To investigate whether the mean polarization is different at 3~GHz compared to 1.4~GHz, we varied $\Pi_0$ and found a marginally better fit for a lower value of ${\rm log_{10}} \Pi_0 = -1.6$, corresponding to $ \bar{f}=0.026$ (reduced $\chi^2=1.3$; probability of $\chi^2$ given the five degrees of freedom of 0.13) (the other parameters were too poorly constrained to optimize), suggesting that the flat-spectrum AGN population that is more prevalent at 3~GHz may be reducing the mean value. Indeed, as shown in Figure \ref{fig:Pol_vs_alpha}, there is a noticeable trend for flat-spectrum ($\alpha > -0.5$) objects to have low fractional polarization that is significant at the 0.1\% level based on the Pearson correlation coefficient. This is consistent with the trend seen by \citet{2002A&A...396..463M} for bright ($>$80~mJy) radio sources in NVSS, where steep-spectrum and flat-spectrum objects have similar polarization fractions at $\approx$1.4~GHz, but where the steep-spectrum sources become more highly-polarized relative to the flat-spectrum population at higher frequencies.
\vspace{-10pt}

\subsection{Source Catalogs}\label{sec3.3}

We include both a Stokes $I$ and a polarized flux component catalog. The catalog columns are detailed in Tables \ref{tab:icat} and \ref{tab:pol_cat}, respectively. 
The catalogs were made with pyBDSF~\citep{2015ascl.soft02007M}, using a 
detection threshold of 5$\sigma$ and an island boundary threshold of 3$\sigma$, with the $\sigma$ map calculated in a 60-pixel box stepped by 30-pixels. The in-band spectral index in the Stokes $I$ catalog is from the Taylor term images (Alpha\_tt) and is estimated by sampling the spectral index image produced by CASA at the peak pixel of the source component, whereas that from the cubes was made using a $\chi^2$ fit to the flux densities of the peak pixels in the channelized images. The catalogs will be made available online (see Data Availability section).

In the polarized source catalog, we give the peak Stokes $I, Q$, and $U$ flux densities in each of the four channels along with polarized flux densities, fractional polarizations, and fitted spectral indices. 
We also include an estimate of $\chi^2$ for the spectral-index fit so objects with potentially curved spectra can be picked out. The uncertainty in the spectral index is calculated by adding in quadrature the error in the fit and a systematic component. This systematic component (0.03) was estimated from the variation in phase calibrator spectral index measured in different execution blocks, and is likely to be dominated by the error in transferring the bandpass calibration from the bandpass calibrator (3C 138) to the phase calibrator and target. 

\begin{table}[H]
\caption{The columns in the Stokes I continuum catalog.}\label{tab:icat}
   {\begin{tabularx}   {\textwidth} {clcL}
   \toprule
    \textbf{Column}
 & \textbf{Quantity} & \textbf{Units} & \textbf{Description}\\\midrule
    (1) & Cmpt & -
 & Component name\\
    (2) & RA   & Degrees & Right Ascension (J2000) \\
    (3) & Unc\_RA  & Degrees (of time) & Uncertainty in R.A.\ \\
    (4) & Dec  & Degrees & Declination (J2000) \\
    (5)&  Unc\_Dec & Degrees& Uncertainty in Dec.\\
    (6) & Total\_flux & mJy & Total flux of fitted Gaussian\\
    (7) & Unc\_Total\_flux & mJy & Uncertainty in Total\_flux\\
    (8) & Peak\_flux & mJy/beam & {{Peak intensity}} of component\\
    (9)& Unc\_Peak\_flux & mJy/beam & Uncertainty in Peak\_flux \\
    (10) & Maj & arcsec & Major axis of fitted Gaussian\\
    (11)& Unc\_Maj & arcsec & Uncertainty in Maj\\
    (12) & Min & arcsec& Minor axis of fitted Gaussian\\
    (13) & Unc\_Min & arcsec & Uncertainty in Min\\
    (14) & PA & Degrees & Position angle (E of N) \mbox{of Gaussian fit}\\
    
      \bottomrule
\end{tabularx}} 
\end{table}

\begin{table}[H]\ContinuedFloat
\caption{\textit{Cont.}}
  {\begin{tabularx}   {\textwidth} {clcL}
   \toprule
    \textbf{Column}
 & \textbf{Quantity} & \textbf{Units} & \textbf{Description}\\\midrule
    
    (15) & Unc\_PA & Degrees & Uncertainty in PA\\
    (16) & DC\_Maj & arcsec & Major axis of fitted Gaussian \\
    & & & deconvolved by the clean beam\\
    (17) & Unc\_DC\_Maj & arcsec & Uncertainty in DC\_Maj\\
    (18)& DC\_Min & arcsec& Minor axis of fitted Gaussian \\ 
    & & & deconvolved by the clean beam\\
    (19) & Unc\_DC\_Min & arcsec & Uncertainty in DC\_Min\\
    (20) & DC\_PA & Degrees & Position angle (E of N)\\
    &&&deconvolved by the clean beam\\
    (21) & Unc\_DC\_PA & Degrees & Uncertainty in DC\_PA \\
    (22) & Isl\_Total\_flux & mJy & Total flux density \mbox{of the source island}\\
    (23) & Unc\_Isl\_Total\_flux & mJy & Uncertainty in Isl\_Total\_flux\\
    (24) & Isl\_rms & mJy/beam & the RMS value of the background\\ &&&around the source island\\
    (25)& S\_Code & - & Source type (S---single, C---a single \\
    &&&Gaussian component, \mbox{with other sources} \\
    &&&on the island, M---a multi-Gaussian source)\\
    (26) & Alpha\_tt & - & In-band spectral index from the\\
    &&&0th and 1st Taylor term images\\
    (27) & Unc\_Alpha\_tt & - & Uncertainty in Alpha\_tt
    \\
\bottomrule
\end{tabularx}}
\end{table}
\vspace{-10pt}
\begin{table}[H]
  \caption{Columns in the catalog of polarized source components.}
    \label{tab:pol_cat}
  {\begin{tabularx}   {\textwidth} {clcL}
  \toprule
 \textbf{Column} & \textbf{Quantity} & \textbf{Units} & \textbf{Description}\\\midrule   
    (1)& Compt &  - & Component name\\
    (2)& R.A. & Degrees & Right Ascension (J2000)\\
    (3)&Unc\_RA  & Degrees (of time) & Uncertainty in R.A. \\
    (4)&Dec  & Degrees & Declination (J2000) \\
    (5)&Unc\_Dec & Degress & Uncertainty in Dec.\\
    (6)&I1 & mJy/beam & Stokes $I$ {{peak intensity}} in channel 1\\
    (7)&Unc\_I1 & mJy/beam & Uncertainty in I1\\
    (8)&Q1 & mJy/beam & Stokes $Q$ peak intensity in channel 1\\
    (9)&Unc\_Q1 & mJy/beam & Uncertainty in Q1\\
    (10)&U1 & mJy/beam & Stokes $U$ peak intensity in channel 1\\
    (11)&Unc\_U1 & mJy/beam & Uncertainty in U1\\
    (12)&I2 & mJy/beam & Stokes $I$ peak intensity in channel 2\\
    (13)&Unc\_I2 & mJy/beam & Uncertainty in I2\\
    (14)&Q2 & mJy/beam & Stokes $Q$ peak intensity in channel 2\\
    (15)&Unc\_Q2 & mJy/beam & Uncertainty in Q2\\
    (16)&U2 & mJy/beam & Stokes $U$ peak intensity in channel 2\\
    (17)&Unc\_U2 & mJy/beam & Uncertainty in U2\\    
    (18)&I3 & mJy/beam & Stokes $I$ peak intensity in channel 3\\
    (19)&Unc\_I3 & mJy/beam & Uncertainty in I3\\
    (20)&Q3 & mJy/beam & Stokes $Q$ peak intensity in channel 3\\
    (21)&Unc\_Q3 & mJy/beam & Uncertainty in Q3\\
    (22)&U3 & mJy/beam & Stokes $U$ peak intensity in channel 3\\
    (23)&Unc\_U3 & mJy/beam & Uncertainty in U3\\   
    (24)&I4 & mJy/beam & Stokes $I$ peak intensity in channel 4\\
    (25)&Unc\_I4 & mJy/beam & Uncertainty in I4\\
    (26)&Q4 & mJy/beam & Stokes $Q$ peak intensity in channel 4\\
    (27)&Unc\_Q4 & mJy/beam & Uncertainty in Q4\\
    
    \bottomrule
\end{tabularx}} 
\end{table}

\begin{table}[H]\ContinuedFloat
\caption{\textit{Cont.}}
{\begin{tabularx}   {\textwidth} {clcL}
  \toprule
 \textbf{Column} & \textbf{Quantity} & \textbf{Units} & \textbf{Description}\\\midrule   
    
    (28)&U4 & mJy/beam & Stokes $U$ peak intensity in channel 4\\
    (29)&Unc\_U4 & mJy/beam & Uncertainty in U4\\
    (30)&Flux & mJy/beam & Peak intensity at 3.0 GHz from \\
    &&&a fit to the channel peak intensities\\
    (31)&Unc\_Flux & mJy/beam & Uncertainty in Flux\\
    (32)&Pol\_flux & mJy/beam & Linearly polarized peak intensity at 3.0~GHz\\
    (33)&Unc\_Pol\_flux & mJy/beam & Uncertainty in Pol\_flux\\
    (34)&Pol\_frac & - & Fractional polarization\\
    (35)&Alpha & - & Spectral index from 2--4 GHz from \\
    &&&a fit to the channel peak intensities\\
    (36)&Unc\_alpha & - & Uncertainty in Alpha\\
    (37)&ChiSq & - & Chi-squared value for the fit\\\bottomrule
  \end{tabularx}}
  \end{table}

\subsection{Polarization Properties of the Brightest Sources in the Sample}\label{sec3.4}

We focused our detailed analysis on the 109 source components  with Stokes $I$ flux densities $>$10~mJy (such that a 5$\sigma$ detection of polarized flux density corresponds to a fractional polarization of $\approx$2.5\%). For each component, Stokes $Q$ and $U$ values were extracted from the cube at the position of the peak pixel in Stokes $I$. 
Estimates of the polarization angle $\phi = 0.5\arctan(U/Q)$ and the fractional polarization were then made. 

We adopted a two-pronged approach to 
analyzing the polarization properties of the bright sources. First, we fit the linear trend of $\phi$ with $\lambda^2$ to determine the rotation measure ($RM$n) of any foreground screen. We then also searched for sources with high levels of depolarization as revealed by a decreasing $\Pi$ with $\lambda^2$.

\begin{figure}[H]
  \hspace{-2.5cm}  \includegraphics[width = 1.0\textwidth]{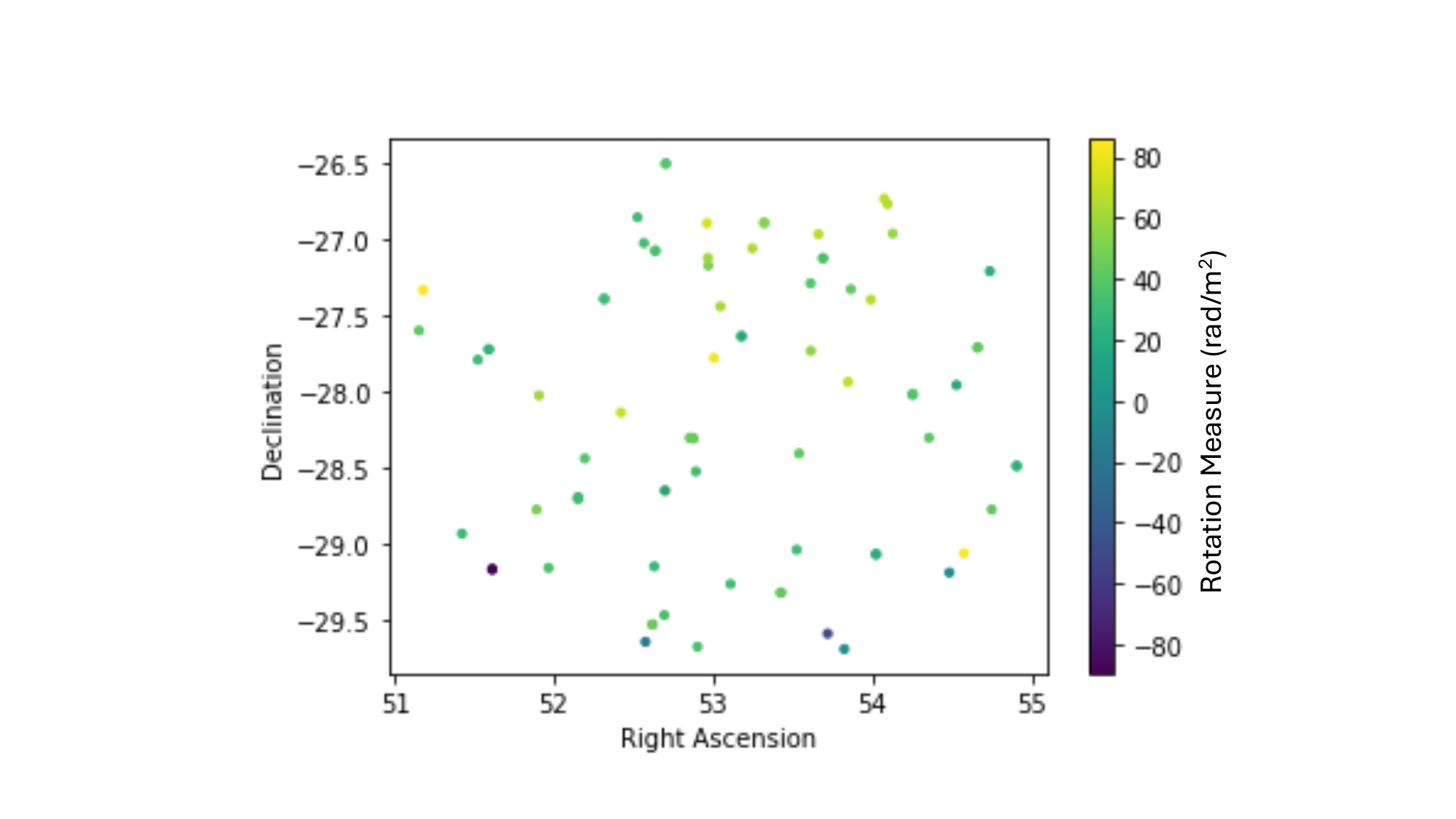}
    \caption{The rotation measure of the brightest sources. The negative values are due to the mean line-of-sight magnetic field being directed away from us.{{The} field is centered at Galactic coordinates $l=224.1, b=-54.6$.}}
    \label{fig: RM map}
\end{figure}

 For 37 of the 109 sources, the values of $\phi$ versus $\lambda^2$ did not show a simple linear trend. In most cases, this was traced to low signal-to-noise resulting in noisy estimates of $\phi$ in each channel, but in some cases, the $RM$ was high and the solution for $\phi$ changed domains in the $Q,U$ plane. An adjustment of 180 degrees was applied to some channels of the latter group, resolving this wrapping problem. The Faraday Rotation and its uncertainty was estimated using linear regression to fit $RM=(\phi-\phi_0)/\lambda^2$, where $\phi_0$ is the original polarization position angle in the absence of Faraday Rotation. These results were plotted two different ways: the first being as a function of sky position as in Figure \ref{fig: RM map}, and then as a histogram in Figure \ref{fig: RM bar graph}. The peak in the histogram at $RM\approx +35$ to $+40~{\rm rad\,m}^{-2}$ is consistent with the distribution of $RM$ in the inner part of the field from ATLAS~\citep{2014MNRAS.441.2555H} and the mean $RM$ value of +34 ${\rm rad\,m}^{-2}$ from the 30 sources of the consolidated catalog of \citet{2023arXiv230516607V} with $50^{\circ}< \mathrm{R.A.}<55^{\circ}$ and $-30^{\circ} < \mathrm{Decl.} < -25^{\circ}$, and is likely to be due to Faraday Rotation from the Milky Way.

 \vspace{-10pt}
\begin{figure}[H]
 \hspace{-0.7cm}   \includegraphics[width = 0.8\textwidth]{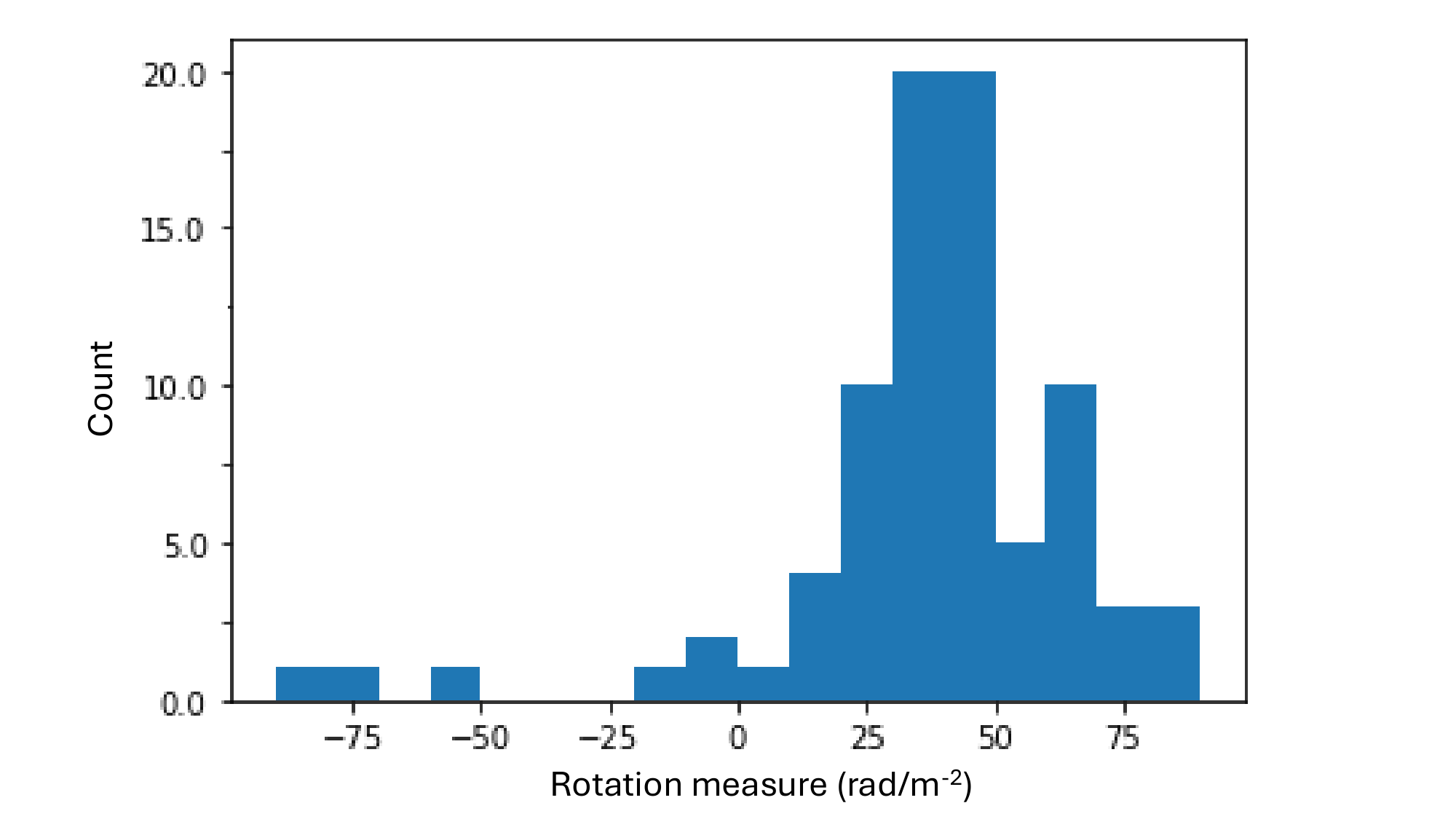}
    \caption{The distribution of rotation measure for the brightest sources.}
    \label{fig: RM bar graph}
\end{figure}

\subsection{Polarization of Faint Sources in the Sample}\label{sec3.5}
Below flux densities of $\sim$1 mJy, the radio source population changes from being dominated by AGN at bright flux densities to being dominated by star-forming galaxies, e.g.,~\citep{Matthews_2021}. We might therefore expect a difference in the mean polarization of the radio source population at flux densities $\stackrel{<}{_{\sim}}$1~mJy compared to that discussed above, depending on the degree of order in magnetic fields in star-forming galaxies on large ($\stackrel{>}{_{\sim}}$1 kpc) scales.

Our survey is not sensitive enough to detect polarization in individual sources at these faint flux-density levels, so we stacked the 
851 sources with $S <$ 1~mJy (median flux density 0.53 mJy) to obtain a limit on the fractional polarization. 
This was achieved by taking \mbox{60 $\times$ 60 pixel} ($90^{''} \times 90^{''}$) polarized intensity sub-images centered around each source component (ignoring the $\stackrel{<}{_{\sim}}$10\% differences in dirty beams as a function of position in the mosaic) and stacking them to create mean polarization image. The process was then repeated for a central position shifted 20 pixels from the source, and difference images were created for the stacks. 

Our initial stacking used the channel-averaged polarization image described in \mbox{Section~\ref{sec:imaging}}. One disadvantage of the above approach is that the high noise in the individual channels dilutes the ability to detect a weak average polarization amplitude signal ($|P|=\sqrt{Q^2+U^2}$ where $P=Q+iU$) even when the channels are averaged together after calculating the $|P|$ image in each channel. \citet{2014ApJ...785...45R} show that the sensitivity of a wideband polarization stack can be improved by applying the mean rotation measure in the field (typically from the Milky Way) to all objects in the stack. This Faraday-rotated image does not help with reducing bandwidth depolarization due to magnetized plasma in the local environments of the sources, but does remove that due to the Milky Way. The Faraday-rotated image is made via Faraday synthesis~\citep{2005A&A...441.1217B}. Defining $\Phi(RM,x,y)= \int P(x,y,\lambda^2)\,e^{2\pi i RM \lambda^2} d(\lambda^2)$ (in our case, approximated by applying the trapezoidal rule over the four channels), we set the $RM$ to the measured average over the field, $RM= +37~ {\rm rad\, m}^{-2}$, and use the $\Phi(37~{\rm rad\, m}^{-2},x,y)$ image for stacking. 

As discussed by \citet{2014ApJ...787...99S}, the Rice distribution of the noise in polarization images means that stacking results in polarization need to be carefully interpreted. Furthermore, at these flux densities, the slope of the radio source counts can mean that the results of stacking are biased by a small number of very bright sources~\citep{2013ApJ...768...37C}. We therefore constructed simulated stacks from both the channel-averaged and Faraday-rotated polarized intensity images using two models for the polarized fluxes: (1) a constant fractional polarization for all objects and (2) objects with polarizations randomly drawn from the BG04  model used in Section~\ref{sec:polcounts}. Our faint source stack showed no signal and few outliers (Figure \ref{fig:polf_faint_diff}), so we used Monte Carlo simulations based on Gaussian noise in $Q$ and $U$ to assess how tight a limit we could place on the polarization of the faint population.

For case (1), fixed fractional polarization, we found that a polarization fraction of $\approx$6.4\% would be detectable in the channel-averaged stack at a 99\% confidence level using the Mann--Whitney $U$ test, or $\approx$3\% in the Faraday-rotated image. In case (2), using the BG04 distribution (Equation \eqref{eqn:polcts}), we found limits of $\Pi_0=4.5$\% and 3.0\% in the channel-averaged and Faraday-rotated image, respectively.
 These limits from stacking are consistent with previous estimates of the polarization of objects of similar flux densities at 1.4~GHz, where \citet{2014ApJ...787...99S} find a limit of $\Pi_0 \stackrel{<}{_{\sim}} 2.5$\% at 2~mJy from stacking NVSS sources and \citet{2014ApJ...785...45R} find $<$2\% polarization for objects with 1.4~GHz 
flux densities between 0.13 and 0.5 mJy.

\vspace{-12pt} 
\begin{figure}[H]
  \hspace{-1cm}  \includegraphics[width = 0.8\textwidth]{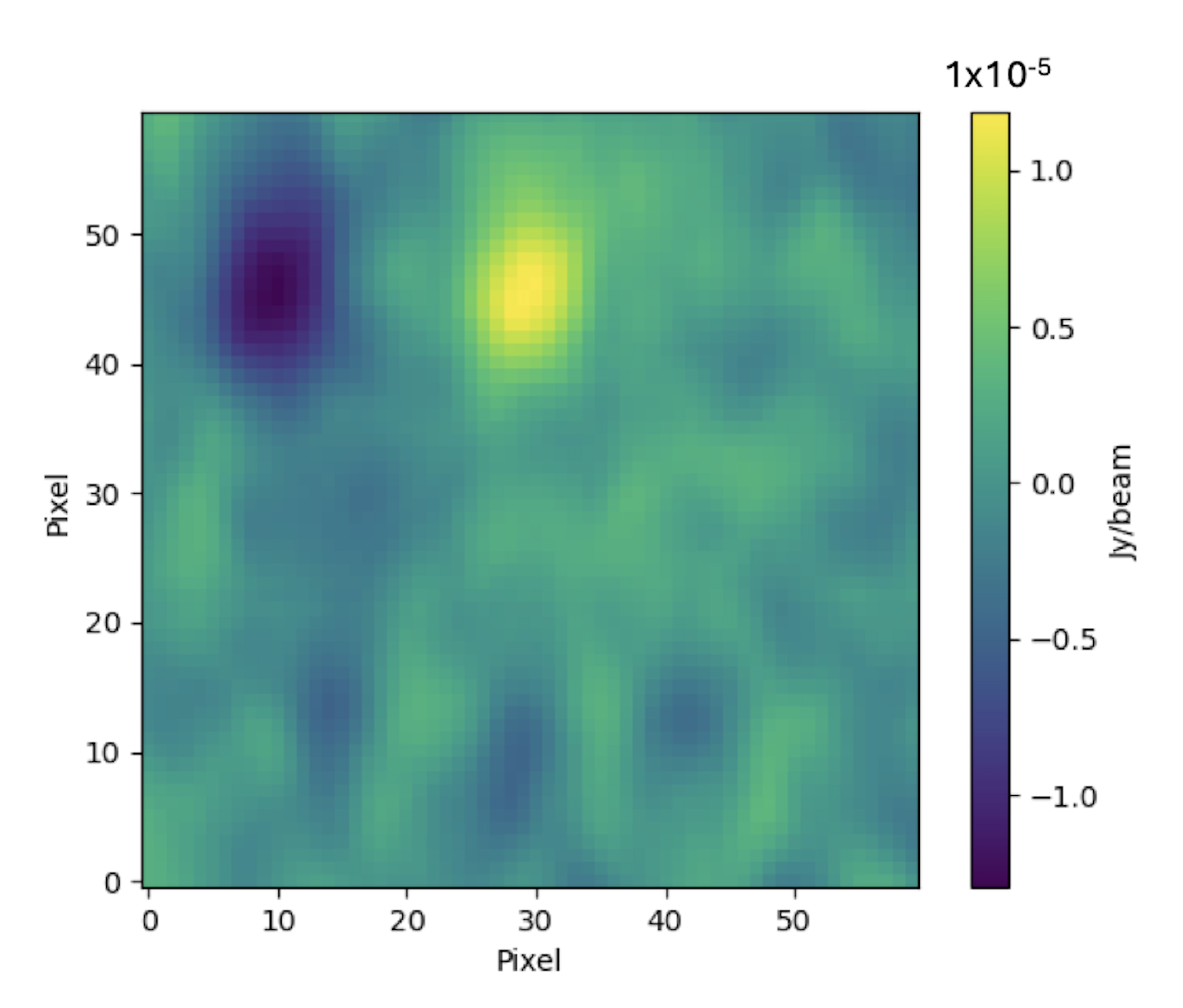}
    
\caption{\textit{Cont.}}
\end{figure}

\begin{figure}[H]\ContinuedFloat

     \hspace{-0.7cm}  \includegraphics[width = 0.7\textwidth]{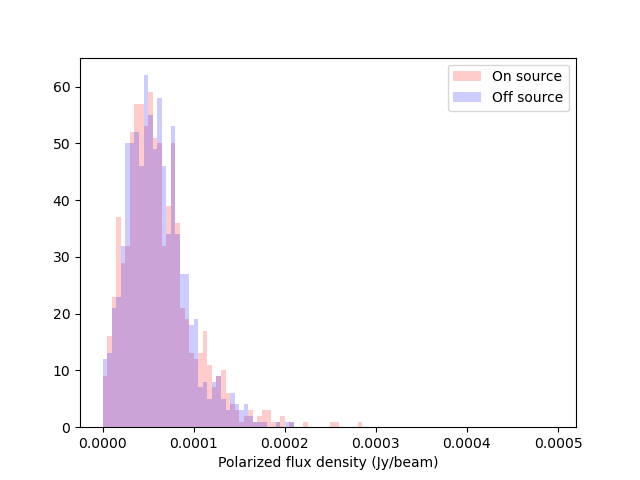}
    \caption{\textbf{Top:} The mean difference between on-source and off-source polarization images. \textbf{Bottom:} The distribution of off- and on-source polarized intensity contributing to the central pixel of the stack.}
    \label{fig:polf_faint_diff}
\end{figure}

\section{Conclusions}\label{sec4}

In this paper, we have provided results from a total intensity and linear polarimetric survey of the W-CDFS field in S-band (2--4~GHz) using the VLA. Using the total intensity (Stokes $I$) data, we show that, once resolved sources in VLASS are taken into account, the 3~GHz source counts as a function of flux density from this survey, SPASS, VLASS, and S-COSMOS are all consistent. 

 Out of the 1920 source components in our Stokes $I$ catalog, 175 have detections of linear polarization. We find that the polarized source counts at 3~GHz are consistent with a mean polarization $\approx$3\%, similar to the population at 1.4~GHz. We also examined the polarization properties of sources in the sample as a function of flux density, by breaking the sample into bright and faint subsamples. The bright subsample contained 109 source components that had total flux densities above 10~mJy.  From these, it was determined that the main source of Faraday Rotation came from the Milky Way, which has a $RM$ of about +35 to +40 rad\,m$^{-2}$ in this part of the sky. 
 (In a future publication (S. Branham et al. in preparation),
 we will examine this subset of highly polarized sources in more detail by studying their polarization properties with rotation measure synthesis~\citep{2005A&A...441.1217B}, as well as their  host galaxies and environments).
The other subset consisted of sources with flux densities below 1~mJy, which we expect to contain a high fraction of star-forming galaxies. The fractional polarization for this group was $\stackrel{<}{_{\sim}}$3\% in the polarization image that had the effect of Milky Way Faraday Rotation removed. 
Comparing to limits $\approx$2--2.5\% at similar flux density levels at 1.4~GHz, these results suggest that there is no large ($>$50\%) difference between the fractional polarization of the radio source population between 1.4~GHz and 3.0~GHz in surveys with similar angular resolution, even though there is $\approx$4$\times$ less Faraday Rotation (and therefore less depolarization) at the higher frequency. This suggests a lack of strongly ordered kpc-scale magnetic fields in star-forming galaxies, though deeper data will be needed to confirm this.

Higher-resolution surveys such as VLASS should result in slightly higher mean fractional polarizations by decreasing the depolarization due to averaging over the beam in extended ($\stackrel{>}{_{\sim}}$3$^{''}$) sources. The large numbers of sources in all-sky surveys such as VLASS will also help with improving the limits from stacking; however, the limit on fractional polarization obtained via stacking is affected strongly by the survey noise, more so than in a conventional stacking experiment, so the stacking limits from VLASS (\mbox{RMS $\approx$ 100 $\upmu$Jy/beam} in the final coadd) may not be much better than those achieved here. Deep, high-resolution surveys are probably best placed to understand the polarization properties of distant star-forming galaxies.





\vspace{6pt}

\authorcontributions{Conceptualization and methodology, S.A. and M.L.; software, S.A., M.L., and P.J.; writing---original draft preparation, S.A. and M.L., writing---review and editing, P.J., J.A., W.N.B., B.M.G., E.H., A.K., J.M., H.M., S.M., R.N., K.N., W.R., N.S., M.V., and R.W. All authors have read and agreed to the published version of the manuscript.}


\funding{This research received no external funding.}

\dataavailability{The component catalogs and Stokes I continuum image are available via Zenodo, DOI: 10.5281/zenodo.18202043. The polarization images are available upon application to the corresponding author, mlacy@nrao.edu.}

\acknowledgments{We would like to acknowledge the National Radio Astronomy Observatory, the National Science Foundation's Research Experience for Undergraduates program, and the University of Rhode Island for their support in the completion of this project. The National Radio Astronomy Observatory is a facility of the National Science Foundation operated under cooperative agreement by Associated Universities, Inc.}

\conflictsofinterest{The authors declare no conflicts of interest.} 





\begin{adjustwidth}{-\extralength}{0cm}

\reftitle{References}



\PublishersNote{}
\end{adjustwidth}
\end{document}